\documentclass[twocolumn,aps,prb]{revtex4}
\usepackage{graphicx,epstopdf,amsmath,amssymb,multirow,CJK,color}

\newcommand {\C}{\textcolor {red}}

\begin{document}

\title{Electronic structures of quasi-one-dimensional cuprate superconductors Ba$_2$CuO$_{3+\delta}$}

\author{Kai Liu$^{1}$}\email{kliu@ruc.edu.cn}
\author{Zhong-Yi Lu$^{1}$}\email{zlu@ruc.edu.cn}
\author{Tao Xiang$^{2,3}$}\email{txiang@iphy.ac.cn}

\affiliation{$^{1}$Department of Physics and Beijing Key Laboratory of Opto-electronic Functional Materials $\&$ Micro-nano Devices, Renmin University of China, Beijing 100872, China}
\affiliation{$^{2}$Institute of Physics, Chinese Academy of Sciences, Beijing 100190, China}
\affiliation{$^{3}$Collaborative Innovation Center of Quantum Matter, Beijing 100190, China}

\date{\today}

\begin{abstract}

  An intact CuO$_2$ plane is widely believed to be a prerequisite for the high-$T_c$ superconductivity in cuprate superconductors.
  However, an exception may exist in the superconducting Ba$_2$CuO$_{3+\delta}$ materials where CuO chains play a more important role.
  From first-principles density functional theory calculations, we have studied the electronic and magnetic structures of Ba$_2$CuO$_{3+\delta}$.
  The stoichiometric Ba$_2$CuO$_3$ and Ba$_2$CuO$_4$ contain quasi-one-dimensional CuO chains and intact two-dimensional CuO$_2$ planes, respectively.
  In comparison with the nonmagnetic metal Ba$_2$CuO$_4$, Ba$_2$CuO$_3$ is found to be an antiferromagnetic (AFM) Mott insulator.
  It possesses a nearest-neighbor intra-chain antiferromagnetic (AFM) coupling and a weak inter-chain interaction, and its lowest unoccupied band and highest occupied band are contributed by Cu 3$d_{b^2-c^2}$-orbital (or $d_{x^2-y^2}$-orbital if we denote the $bc$-plane as the $xy$-plane) and O 2$p$-orbitals, respectively.
  Total energy calculations indicate that the oxygen vacancies in Ba$_2$CuO$_{3+\delta}$ prefer to reside in the planar sites rather than the apical oxygens in the CuO chains, in agreement with the experimental observation.
  Furthermore, we find that the magnetic frustrations or spin fluctuations can be effectively induced by moderate charge doping.
  This suggests that the superconducting pairing in oxygen-enriched Ba$_2$CuO$_{3+\delta}$ or oxygen-deficient Ba$_2$CuO$_{4-\delta}$ is likely to be mainly driven by the AFM fluctuations within CuO chains.
  
\end{abstract}

%\pacs{74.70.Xa, 74.20.Pq, 74.20.Mn}

%\pacs{74.20.Pq, 74.70.Xa, 74.78.-w, 75.70.Ak}
%\item{Condensed Matter Physics}

\maketitle

\section{Introduction}

The interplay between superconductivity and dimensionality is an interesting issue in condensed matter physics.
In the well-known cuprate superconductors~\cite{Bednorz86,Stewart17}, whose superconducting transition temperature is highest at ambient pressure~\cite{Schilling93,ChuCW94}, a common feature is that they all contain  quasi-two-dimensional CuO$_2$ planes~\cite{Dagotto94,Leggett06} from which the superconducting pairing emerges while other layers just serve as a charge reservoir.
Similarly, superconducting condensation is also believed to take place predominately in quasi-two-dimensional Fe$X$ ($X$=As, Se, ...) layers in iron-based superconductors~\cite{sc1,sc2,sc3,sc4,sc5}.
In recent years, superconducting transitions have also been discovered in Bechgaard-salts organic superconductors~\cite{Bechgaard80, Bechgaard81}, molybdenum chalcogenides~\cite{Armici80, Potel80} and pnictides~\cite{RenZA18}, chromium pnictides~\cite{CaoGH15a,CaoGH15b,CaoGH15c,RenZA18PRM,RenZA17a,RenZA17b}, bismuth iodide~\cite{QiYP18}, nickel-bismuth compounds~\cite{XueQK18}, and other quasi-one-dimensional materials.

More recently, high temperature superconductor Ba$_2$CuO$_{4-\delta}$/Ba$_2$CuO$_{3+\delta}$ has been successfully synthesized under high pressure by Jin and coworkers~\cite{JinCQ18}. There are two possibilities regarding the parent compound of these superconductors. 
One possibility is that Ba$_2$CuO$_{4}$ is the parent compound.
High-$T_c$ superconductivity emerges in Ba$_2$CuO$_{4-\delta}$ when some oxygens are removed from Ba$_2$CuO$_{4}$.
Another possibility is that Ba$_2$CuO$_3$ is the parent compound and the superconductivity emerges in Ba$_2$CuO$_{3+\delta}$ when more oxygens are doped to it. 
For the latter, unlike the parent compound of other cuprate superconductors, Ba$_2$CuO$_3$ does not contain CuO$_2$ planes. 
Instead, it contains only CuO chains and the copper-oxygen bond is compressed along the $c$-axis but stretched along the
$b$-axis (the CuO chain direction).

In order to determine the pairing mechanism of electrons, it is crucial to know where the oxygen vacancies reside in the superconducting Ba$_2$CuO$_{4-\delta}$/Ba$_2$CuO$_{3+\delta}$ compounds.
If the oxygen vacancies mainly occupy the apical sites, the CuO$_2$ planes remain intact as the other cuprate superconductors.
Experimental measurements of neutron powder diffraction~\cite{Shimakawa94} and electron diffraction~\cite{Zhang95} for similar compounds, however, suggested that the oxygen vacancies reside mainly on the CuO$_2$ planes and the apical sites are fully occupied by oxygen atoms.
In this case, there are no perfect CuO$_2$ planes.
This implies that intact CuO$_2$ planes are not absolutely necessary in achieving high-$T_c$ superconductivity.

In this paper, we address the above questions by calculating the electronic and magnetic structures of Ba$_2$CuO$_{3+\delta}$ using first-principles density functional theory.
Our study shows that the parent compound is Ba$_2$CuO$_3$, which is an AFM Mott insulator. 
In Sec. I\/I, we first describe the method, and then discuss the numerical results for both stoichiometric Ba$_2$CuO$_3$/Ba$_2$CuO$_4$ and oxygen doped materials. A summary is given in Sec. I\/I\/I.

\section{Results}

\begin{figure}[!t]
\includegraphics[width=0.95\columnwidth]{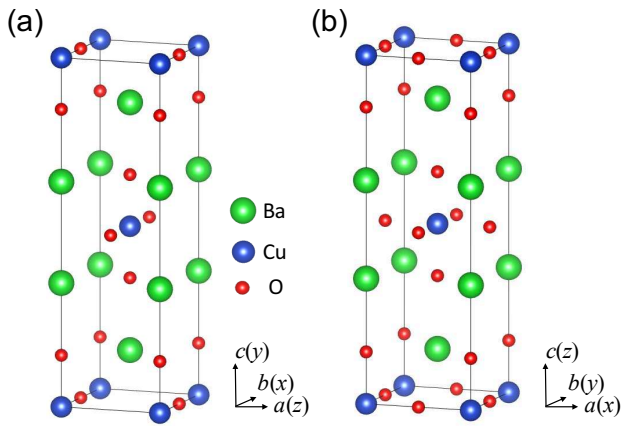}
  \caption{Crystal structures of (a) Ba$_2$CuO$_3$ with CuO chains and (b) Ba$_2$CuO$_4$ with CuO$_2$ planes. Note the $xy$ planes of Ba$_2$CuO$_3$ and Ba$_2$CuO$_4$ are defined differently in order to facilitate the discussion on orbitals.}
\label{fig1}
\end{figure}

We performed the first-principles density functional theory calculations on the electronic and magnetic structures of undoped or doped Ba$_2$CuO$_3$ or Ba$_2$CuO$_4$ materials using the projector augmented wave (PAW) method \cite{PAW} with the Vienna \textit{ab initio} simulation package \cite{VASP1,VASP2}.
The generalized gradient approximation (GGA) of Perdew-Burke-Ernzerhof (PBE) type was used in the construction of the exchange-correlation functional \cite{PBE}.
The plane-wave basis set with a kinetic energy cutoff of 520 eV and a $12\times12\times4$ $k$-point mesh for the Brillouin zone sampling of conventional cell were adopted.
The Fermi level was broadened by the Gaussian smearing method with a width of 0.05 eV.
Ba$_2$CuO$_{3+\delta}$ can be regarded as a compound either by adding oxygen to Ba$_2$CuO$_3$ or by removing oxygen from Ba$_2$CuO$_4$.
Various locations of oxygen vacancies in Ba$_2$CuO$_{3+\delta}$ (or Ba$_2$CuO$_{4-\delta}$) were investigated with both the virtual crystal approximation (VCA) and the supercell methods.
In the latter case, a $2\times2\times1$ supercell with different oxygen vacancy distributions was explored, while the lattice constants were fixed to the experimental values, and all internal atomic positions were allowed to relax until the forces on all atoms were smaller than 0.01 eV/\AA.
The correlation effect among Cu 3$d$ electrons was incorporated using the GGA+U formalism of Dudarev \textit{et al.}~\cite{DFTU} with a typical effective Hubbard interaction U of 6.5 eV.

\subsection{Electronic and magnetic structures of stoichiometric Ba$_2$CuO$_3$ and Ba$_2$CuO$_4$}

Figures \ref{fig1}(a) and \ref{fig1}(b) show the crystal structures of Ba$_2$CuO$_3$ and Ba$_2$CuO$_4$, respectively.
The former is composed of CuO chains along the $b$-axis, while the latter contains CuO$_2$ planes.
In either case, Cu-O layers are separated by Ba-O layers.
Cu atoms in both Ba$_2$CuO$_3$ and Ba$_2$CuO$_4$ form a square-like lattice in each $ab$-plane.
Ba$_2$CuO$_4$ has the same lattice structure as La$_2$CuO$_4$.
However, it is only in Ba$_2$CuO$_3$ that Cu ions have the same nominal valence $+2$ as in La$_2$CuO$_4$.

\begin{figure}[!t]
\includegraphics[width=0.9\columnwidth]{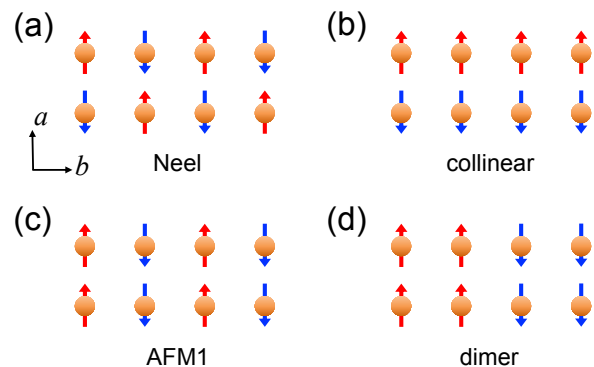}
  \caption{Spin configurations of (a) AFM N\'eel state, (b) collinear AFM state, (c) AFM1 state, and (d) AFM dimer state for the Cu$^{+2}$ spins in the $ab$ plane. The brown balls denote Cu atoms. The red and blue arrows represent the up and down spins, respectively. }
\label{fig2}
\end{figure}

The $a$ and $b$ axes in Ba$_2$CuO$_4$ are equivalent because there are no O vacancies in the CuO$_2$ planes.
However, they are not equivalent in Ba$_2$CuO$_3$ [Fig. \ref{fig1}(a)].
To determine the ground states of Ba$_2$CuO$_3$ and Ba$_2$CuO$_4$, we calculated the energies of several representative magnetic states formed by Cu$^{2+}$ ions and compared with that of the nonmagnetic state.
These magnetic ordered states, as partially shown in Fig.~\ref{fig2}, include the standard AFM N\'eel state, the collinear AFM (or stripe) state, the AFM1 state in which Cu spins are antiferromagnetically coupled along the $b$-axis but ferromagnetically coupled along the $a$-axis, the AFM dimer state~\cite{Gong15, Liu16} along the $b$-axis which is ferromagnetically coupled along the $a$-axis, and the ferromagnetic (FM) state.

For Ba$_2$CuO$_3$, the relative energies of the above magnetic states with respect to the nonmagnetic state are given in Table \ref{TabI}.
All these magnetic states have lower energies than the nonmagnetic one.
Among them, the AFM N\'eel state is energetically degenerate with the AFM1 state, and the collinear AFM state is energetically degenerate with the FM state.
The ground state is found to have either the AFM N\'eel or the AFM1 order, indicating that there is a strong intra-chain AFM coupling along the $b$-axis but a weak inter-chain coupling along the $a$-axis.
The AFM dimer state is next to the ground state in energy.
In contrast, for Ba$_2$CuO$_4$, all the above magnetic states are unstable and the ground state is nonmagnetic.
From the calculations, we find that the weak inter-chain coupling in Ba$_2$CuO$_3$ is about 0.2 meV, smaller than the corresponding values in Ca$_2$CuO$_3$ (3.6 meV) and in Sr$_2$CuO$_3$ (0.8 meV)~\cite{Rosner97,Graaf00}.

\begin{table}[!b]
 \caption{\label{tabI} Relative energies (in unit of meV/Cu) of several magnetic states with respect to the nonmagnetic (NM) state for undoped Ba$_2$CuO$_3$.}
 \renewcommand\arraystretch{1.4}
 \begin{center}
 \begin{tabular*}{0.95\columnwidth}{@{\extracolsep{\fill}}cccccc}
 \hline\hline
 &  N\'eel & collinear & AFM1 & dimer & FM \\
\hline
Ba$_2$CuO$_3$   &  -180  &   -22  &  -180  &  -116  &  -22  \\
\hline\hline
\end{tabular*}
\end{center}
\label{TabI}
\end{table}

\begin{figure}[!t]
  \includegraphics[width=0.85\columnwidth]{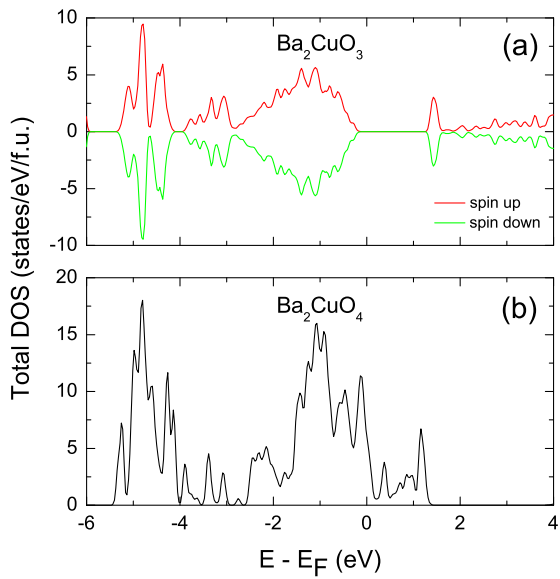}
  \caption{Density of states (DOS) for (a) the  AFM N\'eel state of Ba$_2$CuO$_3$ and (b) the nonmagnetic state of Ba$_2$CuO$_4$. }
  \label{fig3}
\end{figure}

Figure~\ref{fig3} shows the total density of states (DOS) for Ba$_2$CuO$_3$ in the AFM N\'eel state and Ba$_2$CuO$_4$ in the nonmagnetic state, respectively.
In contrast to Ba$_2$CuO$_3$ which is an AFM insulator, Ba$_2$CuO$_4$ is found to be a nonmagnetic metal.
The calculated band gap of Ba$_2$CuO$_3$ is 1.2 eV, close to the experimental results for Ca$_2$CuO$_3$ (1.7 eV) and Sr$_2$CuO$_3$ (1.5 eV)~\cite{Maiti98}.
It should be pointed out that the AFM insulating ground state of Ba$_2$CuO$_3$ is obtained only when the on-site Hubbard interaction is included in the calculation.
Without this interaction, the ground state is metallic.
This suggests that Ba$_2$CuO$_3$ is an AFM Mott insulator~\cite{Schlappa12}, similar to La$_2$CuO$_4$~\cite{Wen06}.

\begin{figure}[!t]
  \includegraphics[width=0.85\columnwidth]{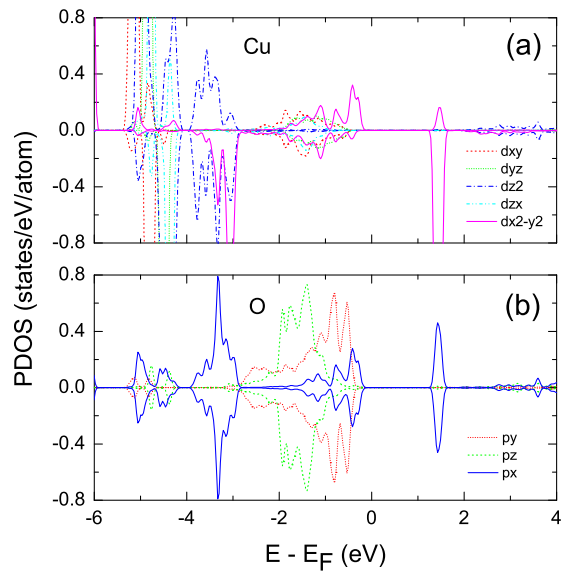}
  \caption{Partial density of states (PDOS) for the AFM N\'eel state of Ba$_2$CuO$_3$ projected on (a) the 3$d$ orbitals of Cu atom and (b) the $2p$ orbitals of O atom along the CuO chain. The up and down parts in each panel represent the spin-up and spin-down channels, respectively. The $xy$ plane is defined in Fig. \ref{fig1}(a).}
\label{fig4}
\end{figure}

\begin{figure}[!b]
  \includegraphics[width=0.8\columnwidth]{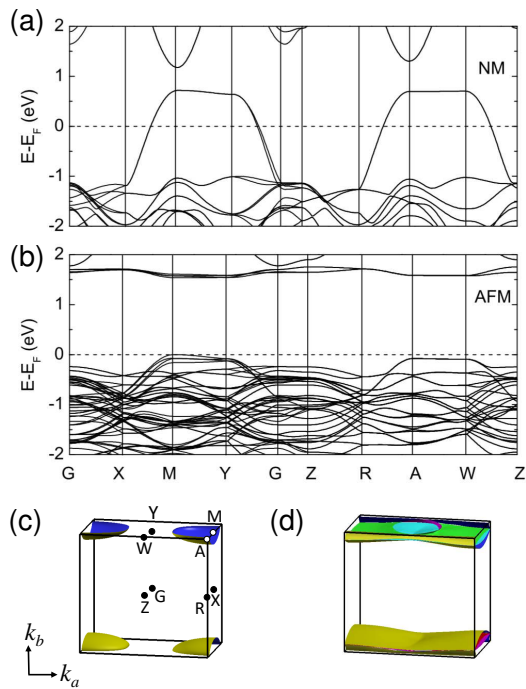}
  \caption{Band structures of BaCuO$_3$ in (a) the nonmagnetic state with a conventional cell and (b) the AFM N\'eel state with a $2\times 2 \times 1$ supercell. (c) and (d) show the Fermi surface contours in the AFM N\'eel state upon hole doping by rigidly shifting the Fermi level 0.05 eV and 0.1 eV below the top of the valence bands, respectively. }
\label{fig5}
\end{figure}

We also calculated the partial density of states (PDOS) for Ba$_2$CuO$_3$ in the  AFM N\'eel state (Fig.~\ref{fig4}).
For convenience, here we denote the $b$ and $c$ axes as the $x$ and $y$ axes [Fig.~\ref{fig1}(a)]. Due to the nominal valence $+2$ of Cu ions in Ba$_2$CuO$_3$, nine electrons fill the 3$d$ orbitals of Cu atom. Among these 3$d$ orbitals, the spin-minority $d_{x^2-y^2}$ orbital is mostly unoccupied [Fig.~\ref{fig4}(a)].
The occupied bands are mainly contributed by O 2$p$ orbitals.
This results from the crystal field effect created by four O atoms surrounding a Cu atom in the $bc$ plane.
The PDOS peaks of unoccupied Cu $d_{x^2-y^2}$ and O 2$p_x$ states appear at almost the same energy above the band gap.
This indicates that there is a strong $p$-$d$ hybridization between these orbitals.

Figures \ref{fig5}(a) and \ref{fig5}(b) show the calculated band structures of Ba$_2$CuO$_3$ along high-symmetry paths of the Brillouin zone in the nonmagnetic and the AFM N\'eel states, respectively.
In the AFM state, the valence band maximum is located at the M point.
Since the CuO chain is along $b$ axis and the inter-chain couplings are rather weak, the band dispersions along the $k_b$ direction are much larger than those along the other two directions.

In oxygen-enriched Ba$_2$CuO$_{3+\delta}$, the extra oxygens doped to Ba$_2$CuO$_3$  introduce hole doping, which is equivalent to shifting the Fermi energy down to the valence bands. In Figs. \ref{fig5}(c) and \ref{fig5}(d), we show the Fermi surface contours when the Fermi level is 0.05 and 0.1 eV below the top of the valence bands in the AFM N\'eel state, respectively.
Upon small hole doping, a hole pocket emerges around the M point.
When more holes are doped into the system, the hole pockets merge together to form flat Fermi surface sheets in the $k_a$-$k_c$ plane.

\subsection{Electronic and magnetic structures of Ba$_2$CuO$_{3+\delta}$ }

In high-$T_c$ cuprates, superconductivity emerges when charge carriers, either holes or electrons, are doped into the parent compounds.
It is commonly believed that AFM spin fluctuations play a very important role in pairing electrons~\cite{Moriya00, Wen06, Scalapino12}.
One can dope holes to Ba$_2$CuO$_3$ either by introducing extra oxygen atoms or by substituting Ba atoms with alkali atoms~\cite{Ueno08,Bollinger11}.
The superconducting material  Ba$_2$CuO$_{3+\delta}$ \cite{JinCQ18,Yamamoto97,Yamamoto00,Karimoto03} can be regarded as a compound either by introducing more oxygens to Ba$_2$CuO$_3$ or by removing oxygens from Ba$_2$CuO$_4$.

To know the electronic and magnetic structures of this material, it is important to know accurately where the oxygen vacancies are located.
We investigated this problem using both the virtual crystal approximation approach and the supercell approach in the framework of density functional theory.

In the virtual crystal approach, we considered three different kinds of distributions of oxygen vacancies: (1) the vacancies are all located in the Cu-O layers, (2) the vacancies are all located in the Ba-O layers, and (3) the vacancies are equally partitioned in both kinds of layers.
For a nominal component Ba$_2$CuO$_{3.2}$, the relative energies of the second and third cases with respect to the first case are found to be +1.88, and +92.58 eV/cell, respectively.
This suggests that the oxygen vacancies prefer to reside in the Cu-O layers.

\begin{figure}[!t]
\includegraphics[width=0.9\columnwidth]{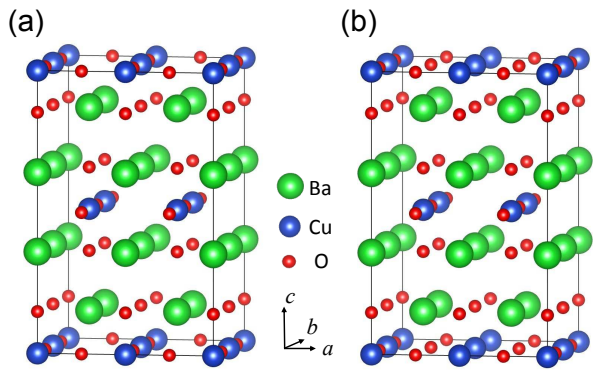}
  \caption{Crystal structures with (a) the lowest energy state and (b) the second-lowest energy state for Ba$_2$CuO$_{3.25}$. They correspond to the 64-1* and 64-1 structures in Table \ref{TabIII}, respectively. }
\label{fig6}
\end{figure}

\begin{table}[!t]
  \caption{\label{tabI} Relative energies (in unit of meV/Cu) of the AFM1 state, the AFM dimer state, and the FM state with respect to that of the nonmagnetic state for Ba$_2$CuO$_3$ under different electron ($e$) and hole ($h$) doping concentrations (per Cu atom).}
\renewcommand\arraystretch{1.4}
\begin{center}
\begin{tabular*}{0.95\columnwidth}{@{\extracolsep{\fill}}cccc}
\hline\hline
doping & AFM1 & dimer & FM  \\
\hline
0          &   -180   &   -116   &   -22   \\
%+0.05   &   -137   &   -90    &   -29   \\
0.1$e$     &   -97     &   -70    &   -35   \\
0.15$e$  &    -61     &   -58    &   -47    \\
%0.2$e$    &   -38     &   -52    &   -61   \\
%-0.05    &   -147   &   -84    &   +4   \\
%0.1$h$      &   -113   &   -57    &   +23   \\
0.15$h$    &   -79     &   -36    &   +11 \\
%0.2$h$      &   -56     &   -20    &   +14   \\
0.3$h$      &   -27     &   -13    &   -3      \\
\hline\hline
\end{tabular*}
\end{center}
\label{TabII}
\end{table}

In the supercell approach, we calculated the energies for 26 possible structures of Ba$_2$CuO$_{3.25}$ (Figs. \ref{fig6} and \ref{fig7}) in a $2\times2\times1$ supercell with 6 oxygen vacancies.
By comparison, we find that the two configurations shown in Figs. \ref{fig6}(a) and \ref{fig6}(b) have the lowest and the second lowest energies, respectively.
In both cases, we also find that the Ba-O layer keeps intact, while the oxygen vacancies also prefer to locate in the Cu-O layers rather than occupying the apical oxygen positions.
The relative total energies of these 26 structures are listed in Table \ref{TabIII}.

The results obtained from both approaches indicate that the oxygen vacancies are located in the Cu-O layers rather than in the Ba-O layers.
This is consistent with the experimental data of neutron powder diffraction~\cite{Shimakawa94} and electron diffraction~\cite{Zhang95} measurements for Sr$_2$CuO$_{3+\delta}$.

To see how the magnetic order is changed by the doping in Ba$_2$CuO$_{3+\delta}$, we evaluated the energies of several different kinds of magnetic states.
Since the inter-chain coupling is very weak, we ignore the difference of different magnetic ordered states along the $a$-axis and assume that the inter-chain coupling along the $a$-axis is ferromagnetic.
Table \ref{TabII} shows how the energies of four typical magnetic states (i.e. nonmagnetic, AFM1, AFM dimer, and FM states) vary with the electron or hole doping.
As expected, the energy difference between the lowest-energy (AFM1) state and the second-lowest-energy (dimer) state is reduced with increasing doping.
This suggests that the doping tends to induce magnetic frustrations~\cite{Liu15} and enhance spin fluctuations, favoring the formation of superconducting Cooper pairs.

\begin{table*}[!t]
  \caption{\label{tabI} Relative total energies (in unit of eV) for all 26 structures of Ba$_2$CuO$_{3.25}$ shown in Figs. \ref{fig6} and \ref{fig7} with oxygen vacancies in a $2\times2\times1$ supercell which contains 16 Ba atoms, 8 Cu atoms, 26 O atoms, and 6 O vacancies.}
\renewcommand\arraystretch{1.4}
\begin{center}
\begin{tabular*}{0.98\textwidth}{@{\extracolsep{\fill}}cccccccccccccc}
\hline\hline
 & BaO-1 & BaO-2 & 82-1 & 82-2 & 82-3 & 73-1 & 73-2 & \C{64-1} & 64-2 & 64-3 & 64-4 & 64-5 & 64-6  \\
\hline
$\Delta{E}$  &  -1.871  &  0  &  -5.168  &  -4.450  &  -4.480  &  -6.350  &  -6.190  &  \C{-8.173}  &  -7.319  &  -7.069  &  -7.614  &  -7.380  &  -7.142  \\
\hline\hline
 & 55-1 & 55-2 & 55-3 & 55-4 & \C{64-1*} & 64-3* & 64-4* & 64-5* & 64-6* & 55-1* & 55-2* & 55-3* & 55-4*  \\
\hline
$\Delta{E}$  &  -7.779  &  -7.736  &  -7.779  &  -7.711  &  \C{-8.201}  &  -7.042  &  -7.606  &  -7.401  &  -7.206  &  -7.729  &  -7.657  &  -7.704  &  -7.683  \\
\hline\hline
\end{tabular*}
\end{center}
\label{TabIII}
\end{table*}

The above discussion suggests that it is Ba$_2$CuO$_3$ rather than Ba$_2$CuO$_4$ that is the parent compound of superconducting Ba$_2$CuO$_{3+\delta}$ materials~\cite{Yamamoto97,Yamamoto00,Karimoto03,JinCQ18}.
This can be more clearly seen by comparison with the other parent compounds of cuprate superconductors, such as La$_2$CuO$_4$.
First, like La$_2$CuO$_4$, Ba$_2$CuO$_3$ is an AFM insulator.
Second, Cu ions in Ba$_2$CuO$_3$ have the same $+2$ nominal valences as in La$_2$CuO$_4$.
Third, the highest occupied and the lowest unoccupied states in both compounds derive from the strongly hybridized O 2$p$ orbitals and Cu 3$d_{x^2-y^2}$ orbitals (for Ba$_2$CuO$_3$, the $bc$ plane is defined as the $xy$ plane and $d_{x^2-y^2}$ is just the $d_{b^2-c^2}$ orbital).

\section{Summary}

We have studied the electronic and magnetic structures of Ba$_2$CuO$_{3+\delta}$ with $\delta$ varying from 0 to 1 by first-principles density functional theory calculations.
Unlike Ba$_2$CuO$_4$ whose ground state is a non-magnetic metal, the ground state of Ba$_2$CuO$_3$ is found to be an quasi-one-dimensional AFM Mott insulator.
The lowest unoccupied and highest occupied states in Ba$_2$CuO$_3$ are mainly contributed by Cu 3$d_{b^2-c^2}$ orbitals and O 2$p$ orbitals, respectively.
By comparison of the total energies of Ba$_2$CuO$_{3+\delta}$ with different kinds of oxygen vacancy structures, we find that the oxygen vacancies reside mainly in the Cu-O layers (planar sites) rather than in the Ba-O layers (apical sites), in agreement with the experimental observation.
Furthermore, doping of charge carriers to Ba$_2$CuO$_3$ can reduce the energy differences between different low-energy magnetic states and thus enhance the spin fluctuations.
This suggests that Ba$_2$CuO$_3$ is the parent compound of oxygen doped Ba$_2$CuO$_{3+\delta}$ superconductors, and the AFM fluctuations in CuO chains play an important role in the superconducting pairing of electrons in these materials.

\begin{acknowledgments}

We thank C. Q. Jin for stimulating discussions. This work was supported by the National Key R$\&$D Program of China (Grant No. 2017YFA0302900) and the National Natural Science Foundation of China (Grants No. 11774422 and No. 11774424). Computational resources were provided by the Physical Laboratory of High Performance Computing at Renmin University of China.

\end{acknowledgments}

\begin{figure*}[tbh]
\begin{center}
\includegraphics[width=0.9\textwidth]{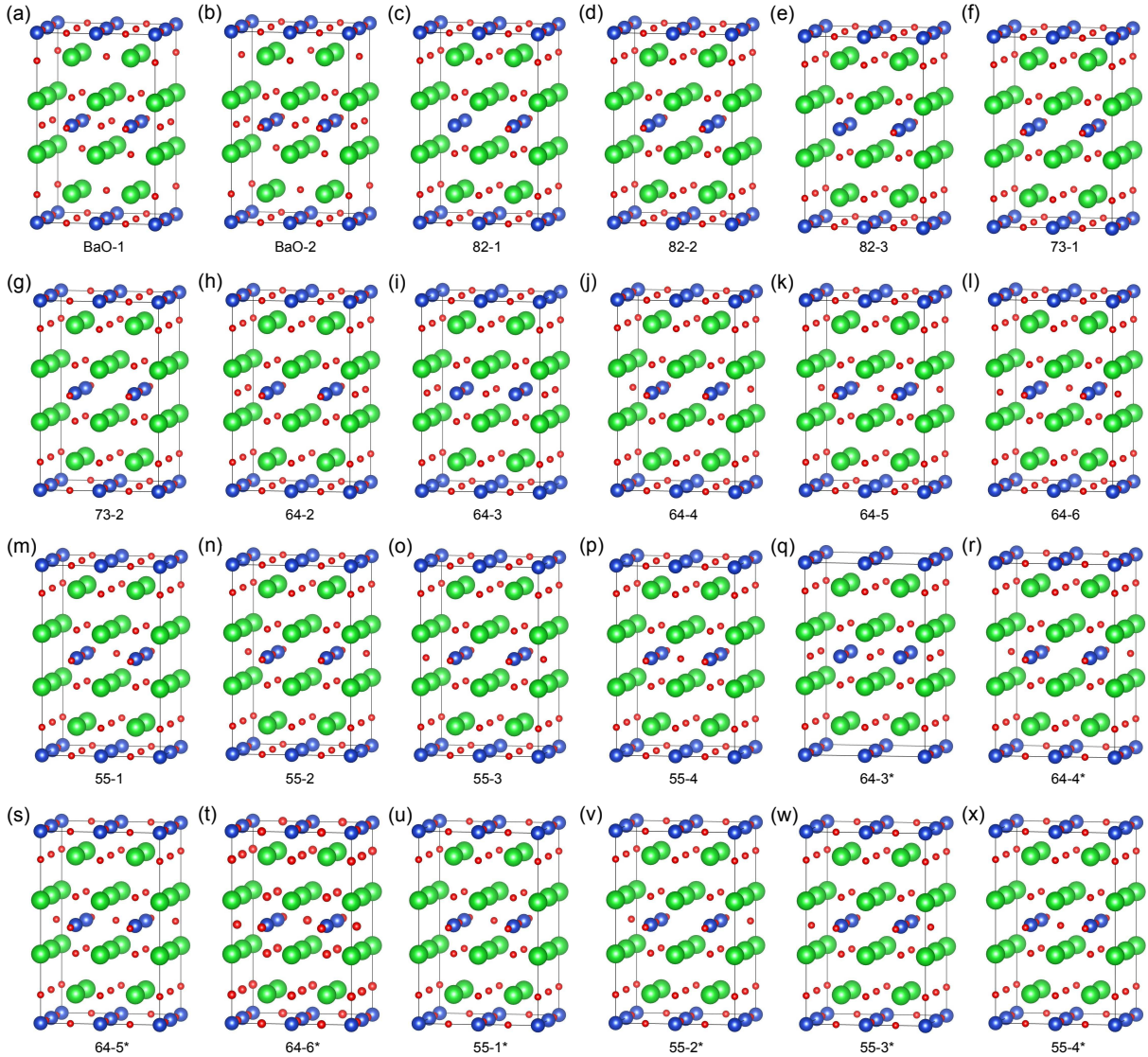}
\end{center}
\renewcommand\arraystretch{1.4}
  \caption{Twenty-four possible structures of Ba$_2$CuO$_{3.25}$. For BaO-1 and BaO-2 structures, O vacancies are in the Ba-O layers. For $mn$-$i$ structures, there are $m$ and $n$ O atoms in the bottom and middle Cu-O layers of the supercell, respectively. The green, blue, and red balls represent the Ba, Cu, and O atoms, respectively.}
\label{fig7}
\end{figure*}

\end{document}